\documentclass[aps,prb,onecolumn]{revtex4}

\usepackage{amsmath}    % need for subequations
\usepackage{amsfonts}
\usepackage{amssymb}
\usepackage{graphicx}   % for figures

\newcommand{\dd}{\mathrm{d}}

\newcommand{\bra  }{\langle}
\newcommand{\ket  }{\rangle}

\usepackage{float}
\usepackage[]{color}
\definecolor{GrisClair}{gray}{0.9}
\begin{document}

\title{Magic composite pulses}
\author{Emmanuel Baudin}
 \affiliation{Laboratoire Kastler Brossel, Ecole Normale Sup\'erieure; CNRS; UPMC; 24 rue Lhomond, F75005 Paris, France\\
 CNRS-Laboratoire de Photonique et de Nanostructures, Route de Nozay, 91460 Marcoussis, France}
 \email{emmanuel.baudin@lpn.cnrs.fr}   %optional
\date{\today}
%\tableofcontents
\begin{abstract}
I describe composite pulses during which the average dipolar interactions within a spin ensemble are controlled while realizing a global rotation. The construction method  used is based on the average Hamiltonian theory and rely on the geometrical properties of the spin-spin dipolar interaction only. I present several such composite pulses robust against standard experimental defects in NRM: static or radio-frequency field miscalibration, fields inhomogeneities. Numerical simulations show that the magic sandwich pulse sequence, a pulse sequence that reverse the average dipolar field while applied, is plagued by defects originating from its short initial and final $\pi/2$ radio-frequency pulses. Using the magic composite pulses instead of $\pi/2$ pulses improves the magic sandwich effect. A numerical test using a classical description of NMR allows to check the validity of the magic composite pulses and estimate their efficiency. 
\end{abstract}

\maketitle

\section{Introduction}

Coupling of spins via the dipole-dipole interaction is of tremendous interest in solid state and liquid state nuclear magnetic resonance\cite{Abragam61} (NMR). It also has deep implications in the field of quantum information processing as dipole-dipole interactions can be used to implement quantum logic gates\cite{Barenco95} but can also produce significative coupling to a thermal bath leading to decoherence\cite{Krojanski06}. 

In solid state NMR, the important strength of the dipolar coupling between first neighbours is an important source of effective transverse relaxation. The ``magic sandwich'' (MS) invented by Rhim, Pines and Waugh \cite{Rhim71} is a pulse sequence that produces an average dipolar interaction equal to minus half of the free dipolar interaction while applied allowing to perform so called time-reversal experiments in samples where the dipolar interactions dominate the Hamiltonian. 

The MS pulse sequence originating from solid state NMR has recently been proven useful in liquid state NMR, namely a hyperpolarized $^3$He-$^4$He liquid mixtures at low temperature\cite{Hayden07} ($\sim 1$~K). In this dense liquid, $^3$He nuclear spins $1/2$ are polarized by optical pumping up to $20\%$ and present a strong long range dipolar interaction \cite{Jeener02} in comparison with solids where the dynamics is dominated by promiscuous spins. Peculiar dynamical behaviours have been observed, in particular spectral clustering at low tip angle and the instability of the uniform magnetization at large tip angle \cite{Nacher00,Sauer01}. In this system, the MS has been successfully applied to stabilize the NMR evolution and should allow to gain insight on the development of incipient turbulent spin dynamics\cite{Hayden07, Lin00}. The MS induced time-reversed evolution is however limited in time, and numerical studies of the NMR dynamics have allowed to point to a built-in defect in the MS pulse sequence due to initial and final $\pi/2$ radio-frequency (rf) pulses which cause a significant perturbation in the average Hamiltonian. 

I have designed and caracterized new composite pulses\cite{Levitt86} having the same qualitative feature as a single $\pi/2$ pulse sequence but being less intrusive on the pure dipolar evolution. Those pulses are based on geometrical constructions derived from the average Hamiltonian theory\cite{Waugh07} and have the same effect as a magic sandwich pulse sequence while applied thus earning their name ``magic composite pulses''. 

This paper is organized in two parts: 
First, we briefly review average Hamiltonian theory applied to the spin-spin dipolar interaction. We extract a geometrical sufficient condition on a pulse sequence to cause an average dipolar Hamiltonian proportionnal to the free dipolar interaction term. A defect of the magic sandwich due to finite pulse length is described and several magic composite pulses are proposed to correct it. 
In the second part, a numerical simulation algorithm allowing the computation of the NMR spin dynamics of classic magnetic moments on a square lattice network is decribed. Numerical simulations allow to evidence the imperfection of the standard magic sandwich. We show that analysis of the spin dynamics computed with this classical model allows to test qualitatively and quantitatively the magic composite pulses. 

\section{Average Hamiltonian theory and construction of the magic composite pulses}
\subsection{Average Hamiltonian}

The spin Hamiltonian in which we will be interested describes the dynamics of like-spins $\hat{\vec {I_i}}$ in a static homogeneous magnetic field submitted to a time-varying rf field and for which we take into account the truncated\cite{Jeener02} dipole-dipole coupling $H_d$. 

Using the convention $\hbar=1$, this Hamiltonian reads

\begin{equation}
H (t) = H_0 + H_{\textrm{rf}} (t) + H_d
\label{eq-Hamiltonian-total}
\end{equation}

where 

\begin{equation}
H_0 = \Omega_0 \sum_i \hat I_{z,i},
\end{equation}

\begin{equation}
H_{\textrm{rf}} (t) = \Omega_1(t) \ \cos{(\omega t + \phi(t))} \sum \hat {I}_{x,i},
\end{equation}

and

\begin{equation}
H_d = \sum_j \sum_{i<j} b_{ij} {}^{t}\hat{\vec {I_i}} \mathbf{A} \hat{\vec {I_j}}
\label{matrice_0}
\end{equation}

with

\begin{equation}
\mathbf{A} =
\left( \begin{array}{ccc}
1&0&0\\
0&1&0\\
0&0&-2
\end{array}
\right) 
\label{matrice_A}
\end{equation}

and 

\begin{equation}
b_{ij} = \gamma^2  \frac{3\cos^2 \theta_{ij} -1}{2   r_{ij}^{3}}
\label{dipolar_field_1}
\end{equation}

In liquid state NMR, this last quantity has an implicit time dependence due to atomic motion. 

If the spin ensemble is described by the density matrix $\rho$, its dynamics is governed by the Liouville equation 

\begin{equation}
\frac{\partial \rho}{\partial t} = -i [H(t),\rho].
\label{eq-Liouville}
\end{equation}

For a static Hamiltonian, a solution of this equation can be simply computed: In this case, the density matrix at time $t_f$ is  related to the one at time $t_0$ by $\rho(t_f) = U(t_f-t_0) \rho(t_0) U^{-1}(t_f-t_0)$ where $U(t)=\exp{(i H t)}$ is the unitary propagator for the time-independent Hamiltonian $H$. 

For a time-dependent Hamiltonian, no simple solution of the Liouville equation exists. 
However, a propagator $U(t_f,t_0)$ can still be defined. 
It is then supposed that this propagator can be written in a form similar to the obtained in the case of a time-independent Hamiltonian: 

\begin{equation}
U(t_f,t_0)=\exp{i S(t_f,t_0)},
\label{eq-Fondamental_Magnus}
\end{equation} 

 where $S$ has the dimensionality of an action\footnote{We recall that we considered $\hbar=1$.}. 
If we know the spin state at time $t_0$ and are only interested in its state at time $t_f$, $\bar{H} = S(t_f,t_0)/(t_f-t_0)$ can be used as effective Hamiltonian instead of $H(t)$ in the Liouville equation\cite{Maricq82}. 

This expression for the propagator $U$ may give wrong intuitions as this average Hamiltonian is only valid to compute the state at time $t_f$ knowing the intial state at time $t_0$. The states obtained at intermediate times have no physical reality and for this reason, the average Hamiltonian method is often described as a stroboscopic method. However, in spite of this inherent limitation it has been widely used with great success in many fields of physics. 

In NMR, the average Hamiltonian method is particularly useful as most NMR pulse sequences are periodic.  In this case, if $\tau$ is the period of the sequence, then the average Hamiltonian used for the time translation from $t_0$ to $t_0 + \tau$ is also valid for any $t_n=t_0 + n \tau$. 
Using symmetry considerations, it is often possible to cancel several orders of the average Hamiltonian. Moreover, in the case of a periodic pulse sequence, Floquet's theorem garranties the existence of the form (\ref{eq-Fondamental_Magnus}) for the propagator $U$. 

\subsection{Magnus expansion of a time-dependent Hamiltonian}

The average Hamiltonian $\bar H$ is a theoretically well defined quantity but pratical determination is by no mean an easy task. Several methods have been described to obtain approximations of $\bar H$. Among them, the most successful in NMR is probably the Magnus expansion \cite{Magnus54} which allows to compute successive aproximations of $\bar H$:

\begin{equation}
\bar H = \bar H^{(0)} + \bar H^{(1)} + \bar H^{(2)} + \cdots
\label{H_limited_development}
\end{equation}

Various methods have been used to express the $\bar H^{(n)}$: first by Magnus \cite{Magnus54} and later by Wilcox\cite{Wilcox67} and Haerberlen\cite{Haeberlen68}. For reference the first two terms are given here considering, without lack of generality, that $t_0=0$ and $t_f=\tau$: 

\begin{equation}
\bar H^{(0)} = \frac 1 \tau \int_0^\tau \dd {t_1} H(t_1),
\label{H0_limited_development}
\end{equation}

\begin{equation}
\bar H^{(1)} = \frac {-i}{2 \tau}\int_0^\tau \dd {t_1} \int_0^{t_1} \dd {t_2} \left[ H(t_1), H(t_2) \right].
\label{H1_limited_development}
\end{equation}

General expression for the term of order $n$ implies $n$ nested commutators of the time-dependent Hamiltonian $H(t)$ at different times. \cite{Oteo00} 

A sufficient condition for this series expansion to converge toward the average Hamiltonian is that the form (\ref{eq-Fondamental_Magnus}) for the propagator exists. 
In the case of a periodic pulse sequence existence, and thus convergence, is garrantied by Floquet's theorem\cite{Maricq82}. The problem for using the magnus expansion is thus rather whether truncation of the Magnus expansion is representative of the sum. This criterion is given by
\begin{equation}
\int_0^\tau \| H(t)\|  \dd t <\pi,
\label{eq_convergence_criterion}
\end{equation}
where $\| . \|$ is the supremum norm:
\begin{equation}
\| A\| = \sup_{\bra x | x \ket \neq 0} \frac{\bra x |A | x \ket}{\bra x | x \ket}.
\label{eq_supremum_norm}
\end{equation}
There are many cases however where this criterion is not met experimentally and where truncation of the Magnus expansion still provides accurate prediction on the experimental observations. The reason for it is that, in a large sample, only few modes of the average Hamiltonian are resonant with the pulse sequence spectrum, so that truncation fails only for these modes but most of the sample is correctly described by the truncation. In this case, the truncation of the Magnus expansion does not predict slow relaxation due to warming of the spins by the rf pulse sequence. This explains why, also in many cases assumption (\ref{eq_convergence_criterion}) is not verified, truncation of the Magnus expansion is used with success in solid state NMR. 

In the case of magic composite pulses, the pulse sequence not beeing periodic, assumption (\ref{eq-Fondamental_Magnus}) is not garrantied. Casas \textit{et al.} have shown that criterion (\ref{eq_convergence_criterion}) is sufficient to have the existence of (\ref{eq-Fondamental_Magnus}) and thus the convergence and validity of the truncation for the Magnus expansion\cite{Casas07,Blanes09}. 
However, in the case of a dipole-dipole Hamiltonian, criterion (\ref{eq_convergence_criterion}) is experimmentally tremendously difficult to meet due to large eigenvalues in the Hamiltonian. 

In fact partial validity of the Magnus expansion beyond criterion (\ref{eq_convergence_criterion}) is provided by the same argument used in the case of Magnus expansion truncation in the case of periodic Hamiltonians: In the general case, the form (\ref{eq-Fondamental_Magnus}) does not exist for the whole system. However, it is possible to define a large subspace of the Hilbert space in which truncation of the Magnus expansion can be used to described the NMR evolution (only few excluded modes beeing warmed by the rf pulse sequence). This imply that the pulse sequence decribed in this paper have to be taken with the same cautiousness as regular periodic magic pulse sequences in solid state NMR as they rely on similar assumptions. 

In the case of quantum information processing, approximate validity arguments probably fail considerably and condition (\ref{eq_convergence_criterion}) must be met absolutely if magic composite pulses are used. 

For a given Hamiltonian, a change in the reference frame can change the domain of convergence of the Magnus expansion. 
A carefully chosen reference frame can also provide faster convergence of the Magnus series\cite{Maricq82}. 
In the case of the dipolar interaction in NMR, the proper reference frame to work with is the doubly tilted rotating frame. \cite{Rhim73, Rhim74, Haeberlen68, Redfield55} 

\subsection{The doubly tilted rotating frame}

The doubly tilted rotating frame is the frame in which the instantaneous Hamiltonian is only the dipolar interaction\cite{Redfield55}. The (passive) rotation operator transforming the laboratory frame in the doubly tilted rotating frame is the propagator computed by taking into account the external and applied magnetic field only. It can be deduced from the Liouville equation and satisfies the equation :

\begin{equation}
i \partial_t R = (H_0 + H_{\textrm{rf}}(t)) R
\end{equation}

with $R(0)=\textrm{Id}$ (the instant $t=0$ is chosen appropriately). 
The Hamiltonian in (\ref{eq-Liouville}) can be replaced by $H'(t)$ in the doubly tilted rotating frame where $H'(t)$ is given by
\begin{equation}
H'= R(t) H R^{-1}(t) - i \partial_t R(t) R^{-1}(t). 
\end{equation}
Using this transformation on the spin Hamiltonian (\ref{eq-Hamiltonian-total}), we get
\begin{equation}
H' (t) = H_d' (t)= R(t) H_d R^{-1}(t),
\label{eq-doubly-tilted}
\end{equation}

as extra terms corresponding to the applied field have been canceled by going into the doubly tilted rotating frame. 

We replace $H_d$ by its expression given in equation (\ref{matrice_0}) and apply the following simplification :
\begin{equation}
R {}^{t}\hat{\vec {I_i}} \mathbf{A} \hat{\vec {I_j}} R^{-1} = (R {}^{t}\hat{\vec {I_i}} R^{-1}) (R \mathbf{A} R^{-1}) (R \hat{\vec {I_j}} R^{-1})
\label{matrice_2}
\end{equation}

at this point we remark that $R(t)$ is an evolution operator which is composed of infinitesimal rotations in the SU(2) space and is consequently a rotation operator this space. 
We call $\mathbf{R}$ the corresponding rotation in SO(3). 
So that we get

\begin{equation}
R \hat{\vec {I_j}} R^{-1}= (\mathbf{R} \hat{\vec {I_j}}),
\label{transformation_1}
\end{equation}

and

\begin{equation}
R {}^{t}\hat{\vec {I_i}} R^{-1} = {}^{t}(\mathbf{R}\hat{\vec {I_i}}).
\label{transformation_2}
\end{equation}

Moreover since $R$ alone is a scalar quantity in the SU(2) space :

\begin{equation}
R \mathbf{A} R^{-1}=\mathbf{A},
\label{transformation_3}
\end{equation}

hence

\begin{equation}
R {}^{t}\hat{\vec {I_i}} \mathbf{A} \hat{\vec {I_j}} R^{-1} = 
{}^{t}(\mathbf{R}\hat{\vec {I_i}}) \mathbf{A} (\mathbf{R} \hat{\vec {I_j}}) =
{}^{t}\hat{\vec {I_i}} \mathbf{A'}(t) \hat{\vec {I_j}}
\label{matrice_3}
\end{equation}

with

\begin{equation}
\mathbf{A'} (t) = \mathbf{R^{-1}}
\left( \begin{array}{ccc}
1&0&0\\
0&1&0\\
0&0&-2
\end{array}
\right) \mathbf{R}.
\label{matrice_A_prime} 
\end{equation}

Finally, we obtain the following expression for $H'_d (t)$ which we will use whenever we apply the Magnus expansion: 

\begin{equation}
H'_d (t) = \sum_j \sum_{i<j} b_{ij} {}^{t}\hat{\vec {I_i}} \mathbf{A'}(t) \hat{\vec {I_j}}
\label{matrice_t}
\end{equation}

Let us remark that equation (\ref{matrice_A_prime}) implies that the matrix $\mathbf{A'} (t)$ can be written in the following simple form

\begin{equation}
\mathbf{A'} (t) = \textrm{Id}-3 {}^{t}\hat{n}(t)\cdot \hat{n}(t)
\label{matrice_A_prime2} 
\end{equation}
where $\hat{n}(t)$ is the unitary vector pointing to the $\hat z$ direction of the laboratory frame from the doubly tilted rotating frame. It is thus solely determine by 2 angles. 

\subsection{Magic conditions}

The computation of $\bar {H}'^{(0)}_d$ is \textit{a priori} complicated, but we remark that the sequence operator $R(t)$ only concerns the vectorial part of the Hamiltonian (\ref{matrice_0}) and does nothing to its spatial dependencies (\ref{dipolar_field_1}). The term $\bar {H}'^{(0)}_d$ is then 

\begin{equation}
\bar{H}'^{(0)}_d = \left[ \sum_j \sum_{i<j} b_{ij} \right] \frac 1 {\tau} \int_0^{\tau} {}^{t}\hat{\vec {I_i}} \mathbf{A'}(t_1) \hat{\vec {I_j}}\  \dd {t_1}.
\label{Hamiltonian_eff0_2}
\end{equation}

We define a magic sequence as a sequence which rotates the spins and does not modify the dipolar interaction  during a given duration (except by a scaling factor). 
The zero order condition for magic sequences formally reads

\begin{equation}
\bar{H}'^{(0)}_d = k H_d.
\label{criterion_0}
\end{equation}
It can easily be shown that $-1/2\leq k \leq 1$.

From this condition we can extract a general geometrical condition by separating the spatial part from the geometrical part of the dipolar interaction. We get the following geometrical condition: 

\begin{equation}
\frac 1 {\tau} \int_0^{\tau} \mathbf{A'} (t_1)  \dd {t_1}= k \mathbf{A}
\label{criterion_1}
\end{equation}

It can be simply verified that conditions (\ref{criterion_0}) and (\ref{criterion_1}) are equivalent. 

For higher orders the condition of order $n$ for a magic sequence is simply: 
\begin{equation}
\bar{H}'^{(n)}_d = 0 \quad \mbox{if $n \geq 1$}
\label{criterion_2}
\end{equation}

As for order 0, we show in appendix \ref{appendix-geometry} that geometrical conditions on the pulse sequence exist and are sufficient to ensure that (\ref{criterion_2}) is verified. The equivalence of those geometrical conditions with the corresponding conditions on the average dipolar spin Hamiltonian is not prooven in the general case (although appendix \ref{appendix-equivalence} equivalence for orders 0 and 1 is prooven). 

\subsection{Taking into account experimental defects}

Experimental NMR pulse sequences are plagued with defects. 
The idea to deal with it consists in separating the time-dependent Hamiltonian (\ref{eq-Hamiltonian-total}) in a ideal part and a perturbative part :

\begin{equation}
H (t) = H_0 + H_{\textrm{rf}} (t) + H_d + \delta H_0 + \delta H_{\textrm{rf}} (t)
\label{Hamiltonian-complet}
\end{equation}

Here $H_0$ is the ideal static field being equal to $\omega I_z$ and $\delta H_0$ is the difference with the applied field. 
Similarly $H_{\textrm{rf}} (t)$ designates the ideal pulse sequence following the theoretician expectations and $\delta H_{\textrm{rf}} (t)$, the errors in the experimental realization. 
We now express Hamiltonian (\ref{Hamiltonian-complet}) in the doubly tilted rotating frame:

\begin{equation}
H' (t)= H_d' (t) + R(t) \delta H_0 R^{-1}(t) + R(t) \delta H_{\textrm{rf}} (t) R^{-1}(t)
\end{equation}

Applying the average Hamiltonian theory we get an intricate average Hamiltonian $\bar{H}'$. 
However at the lowest order, the average Hamiltonian is a simple sum of the three different effects:

\begin{equation}
\bar{H}'^{(0)}= \bar{H}'^{(0)}_d + \bar{\delta H}'^{(0)}_0 + \bar{\delta H}'^{(0)}_{\textrm{rf}}
\end{equation}

\subsubsection{Robustness against a Larmor shift}

If the local Larmor frequency is not equal to the rf, the perturbation brought on the spin dynamics has to be taken into account. 
The extra term in the average Hamiltonian due to a Larmor shift can be easily calculated using equation (\ref{H0_limited_development}) with the Zeeman term $\delta H_0 = \delta \Omega_0 \ \hat{\vec I} \cdot \hat z$. 
We obtain the effective hamiltonian term :

\begin{equation}
\bar{\delta H_0}' = 
\frac 1 {\tau} \int_0^{\tau} R(t) \delta H_0 R(t)^{-1} \dd {t} =
\frac 1 {\tau} \int_0^{\tau} \dd {t} \ \delta \Omega_0 (R(t_1) (\hat{\vec I} \cdot \hat z)  R(t_1)^{-1}) 
\label{Larmor_1}
\end{equation}

We use the same simplification technique :

\begin{equation}
R(\hat{\vec I} \cdot \hat z)  R^{-1}= 
(R\hat{\vec I} R^{-1})  \cdot \hat z=
(\mathbf{R} \hat{\vec I})  \cdot \hat z=
\hat{\vec I} \cdot (\mathbf{R^{-1}} \hat z)
\label{Larmor_2}
\end{equation}

This last expression allows to write a simple geometrical criterion for a sequence to be robust against a Larmor shift: 

\begin{equation}
 \int_0^{\tau} \mathbf{R^{-1}}(t) \hat z \ \dd t = 0
\label{Larmor_3}
\end{equation}

This criterion can be verified easily for a given pulse sequence. 
In NMR experiments, composite pulses robust against a Larmor shift are desirable to prevent failure of the pulse sequence due to experimental imperfections of the field homogeneity or the application of magnetic field gradients. 

\subsubsection{Robustness against perturbation in the excitation field $B_1$}

In the case of a perturbation in the excitation field $B_1$, the condition of robustness at the lowest order is

\begin{equation}
\bar{\delta H}'^{(0)}_{\textrm{rf}} = \frac 1 {\tau} \int_0^{\tau} R(t) \delta H_{\textrm{rf}}(t) R(t)^{-1} \dd {t} = 0.
\end{equation}

Unfortunately in this case no obvious geometrical equivalent condition exists because of the time dependence of the perturbative term $\delta H_{\textrm{rf}}(t)$ which depends on the pulse sequence used. Moreover, if solutions exist for a miscalibration of $B_1$ amplitude, no possibility can be found in the case of a miscalibration of phases (except for pulse sequences doing a $2\pi$ rotation).  

\subsection{Examples of magic sequences of order zero}

\subsubsection{The magic sandwich}

The MS\cite{Rhim71} consists in applying a strong rf between to perfect $\pi/2$ pulses. Here we compute the effective Hamiltonian $\bar{H}^{(0)}$ during the continuous rf field along the $\hat x$-axis. 
The rotation operator $\mathbf{R_x}(t)$ is\footnote{We take the convention that $\Omega_1$ is the rf field amplitude before the rotating wave approximation.}

\begin{equation}
\mathbf{R_x}(t) = 
\left( \begin{array}{ccc}
1&0&0\\
0&\cos \frac {\Omega_1 t}{2}&\sin \frac {\Omega_1 t}{2}\\
0&-\sin \frac {\Omega_1 t}{2}&\cos \frac {\Omega_1 t}{2}
\end{array}
\right),
\label{magic_sandwich_0}
\end{equation}

so 

\begin{equation}
\mathbf{A'} (t) =
\left( \begin{array}{ccc}
1&0&0\\
0&-\frac 1 2 + \frac 3 2 \cos \Omega_1 t& \frac 3 2 \sin \Omega_1 t\\
0& \frac 3 2 \sin \Omega_1 t&- \frac 1 2 - \frac 3 2 \cos \Omega_1 t
\end{array}
\right). 
\label{magic_sandwich_1}
\end{equation}

If we do a $\pi$ rotation the average interaction matrix $\mathbf{\bar A'}$ is

\begin{equation}
\mathbf{\bar A'} =
\left( \begin{array}{ccc}
1&0&0\\
0&-\frac 1 2 &0\\
0& 0&- \frac 1 2
\end{array}
\right). 
\label{magic_sandwich_2}
\end{equation}

$\mathbf{\bar A'}$ is not directly proportional to $\mathbf{A}$, but is  with $k=-1/2$ when sandwiched between two $\pi/2$ rotations along the $\hat y$-axis (this structure is the reason for naming this pulse sequence "magic sandwich"). 

\subsubsection{Effect of the $\pi/2$ pulses finite duration in the MS}

If we suppose that pulses are not infinitely short, we have to take into account the imperfections brought by the initial and final $\pi/2$. 
The initial $\pi/2$ consists of a rotation along the $\hat y$-axis. We compute the effective interaction (the pulse last $\tau = \pi / \Omega_1$) :

\begin{equation}
\frac 1 {\tau} \int_0^{\tau} \mathbf{A'} (t) \dd t=
\left( \begin{array}{ccc}
-\frac 1 2 &0&\frac 3 \pi\\
0&1&0 \\
 \frac 3 \pi&0&- \frac 1 2 
\end{array}
\right) 
\label{normal_half_pi}
\end{equation}

This interaction matrix is not proportional to the free interaction matrix because of those perturbative non-diagonal terms. Consequently, the first and last $\pi/2$ pulses finite duration of the magic sandwich are sources of perturbation on the ideal NMR evolution. An illustration of the effect of those pulses by numerical simulation is given in section \ref{exponential}. 

\subsubsection{A $\pi/2$ magic composite pulse}
\label{ref-Mpis2}
In order to suppress the perturbation introduced by the $\pi/2$ pulses in the MS sequence, we replace the simple $\pi/2$ rotation by the sequence
\begin{equation}
 (\pi_x - (\pi/2)_y - (\pi/2)_{y,1/2} - (\pi/2)_{-y,1/2}). 
\end{equation}
The indexes $1/2$ indicate that the pulses durations are the half duration of the standard pulses. 
The total duration of rf pulses along $\hat x$ is equal to the total duration of rf pulses along $\hat y$. We compute the average interaction matrix by dividing the computation in two terms: the first part consisting of a $\hat x$ rotation and the second part consisting of a $\hat y$ rotation :

\begin{equation}
\mathbf{\bar A'} = \frac 1 2 (\mathbf{\bar A'_x}+\mathbf{\bar A'_y})
\label{MHP_0}
\end{equation}

The geometrical term $\mathbf{\bar A'_x}$ corresponding to the $\pi_x$ pulse is given by equation (\ref{magic_sandwich_2}). The rf along the $\hat y$ axis is more complicated, but it has been shaped so that the time spent in every direction on the circle described on the Bloch sphere is the same and consequently, at order 0, causes the effect of a $2\pi$ homogeneous rotation\footnote{See appendix \ref{appendix-fruitful_transformation} for the occupation density explanation.}.

\begin{equation}
\mathbf{\bar A'_y} = 
\left( \begin{array}{ccc}
-\frac 1 2&0&0\\
0&1&0\\
0& 0&- \frac 1 2
\end{array}
\right) 
\label{MHP_4}
\end{equation}

This concludes the calculation on the effect of the magic pulse sequence: 

\begin{equation}
\mathbf{\bar A'} = 
\left( \begin{array}{ccc}
\frac 1 4&0&0\\
0&\frac 1 4&0\\
0& 0&- \frac 1 2
\end{array}
\right) 
\label{MHP_5}
\end{equation}

the effective dipolar interaction at the order zero is $k=1/4$ and the equivalent rotation during the pulse is $\pi_x$ - $(\pi/2)_y$ which brings the $\hat z$ component of the spin along the $\hat x$ direction. This property makes this pulse sequence suitable to replace $\pi/2$ initial and final pulses in a magic sandwich.

\subsubsection{Other magic composite pulses $\pi/2$}

Using the technique described in the previous section, we can build other M$\pi/2$ which can be more efficient in certain situations. I give few of them here (proofs are left to the reader)

The first inhomogeneous rotation along the $\hat x$-axis can be chosen between:

\begin{equation}
\begin{tabular}{c}
  $\pi_x,\pi_{-x}$ \\
  $2 \pi_x,2 \pi_{-x}$ \\
  $2 \pi_x,2 \pi_{-x}$ \\
  $2 \pi_{\pm x}$ \\
  $\pi_{\pm x}$ 
\end{tabular}
\label{eq-MCP_premiere_partie}
\end{equation}

The second inhomogeneous rotation along the $\hat y$-axis can be chosen between the following sequences:

\begin{equation}
\begin{tabular}{c}
  $\pi/2_{- y,1/2},\pi/2_{y,1/2},\pi/2_{y}$ \\
  $\pi/2_{y},\pi/2_{y,1/2},\pi/2_{-y,1/2}$ \\
  $\alpha_{-y,1/2},\alpha_{y,1/2},\pi/2_{y},(\pi/2-\alpha)_{y,1/2},(\pi/2-\alpha)_{-y,1/2}$ 
\end{tabular}
\label{eq-MCP_deuxieme_partie}
\end{equation}

Those last three pulse sequences have been shaped to give an average dipolar matrix equal to the one obtained using a single $2\pi_y$ pulse. 
By assembling two inhomogeneous rotations among (\ref{eq-MCP_premiere_partie}) and (\ref{eq-MCP_deuxieme_partie}) having the same duration along the $\hat x$-axis and $\hat y$-axis, a magic composite $\pi/2$ pulse is formed.

Several variations are possible, since we only chose one particular strategy. For example, we could decide to mix $\hat x$ and $\hat y$ rotations in a complex way, or use a free evolution period to compensate for diagonal terms. Here is an example where the second part of the sequence is only composed of pulses of the same intensity: 
\begin{equation}
\pi_{x,7/6},\pi/6_{-y},\pi/6_{y},\pi/2_{y},\pi/6_{y},\pi/6_{-y}
\label{MHP_13}
\end{equation}
This sequence has the advantage of being shorter (for the same duration reference) than the first M$\pi/2$ and is then expected to be more efficient. This time, the $\hat y$ sequence is not a simple equivalent to a $2 \pi$ homogeneous rotation, although it has the same properties (see appendix \ref{appendix-fruitful_transformation}). 

Other magic composite pulses robust against a Larmor shift can be created by using the extra condition (\ref{Larmor_3}). 
The following sequence is both magic and robust against a Larmor shift  and realize a total rotation equivalent to $\pi/2_{-y}$ pulse:

\begin{equation}
2 \pi_{x},3\pi/2_{y},\pi/2_{y,0.5},-\pi/2_{y,0.5}
\label{MHP_14}
\end{equation}

All the magic composite $\pi/2$ pulses presented until now are of order 0. In appendix \ref{appendix-Mpulse1}, I describe the construction of a M$\pi/2$ of order 1 :

\begin{equation}
\begin{split}
[(\pi,-\pi)_{y,2.767},(\pi,-\pi)_{x,2.767},(\pi,-\pi)_{x,2.767}],[\pi_{y,1.488},
(\pi/4_{1.557},\pi/4_{1.},\pi/4_{2.557})_y,\\
(\pi/4_{2.557},\pi/4_{1.},\pi/4_{1.557})_y,\pi_{y,1.488}]
\end{split}
\end{equation}

This pulse sequence is particularly long and difficult to realize with an NMR spectrometer but will nontheless be used in numerical simulations to illustrate the efficiency of a numerical test for magic composite pulses. 

\subsection{Enhanced magic sandwich}
\label{partie-EMS}

Using magic composite $\pi/2$ it is possible to enhance the efficiency of the MS (see section \ref{exponential}). The usual MS pulse sequence is 

\begin{equation}
\pi/2_y, (2 \pi_x, -2 \pi_x)_n, \pi/2_y,
\end{equation}
where the filling with $(2 \pi, -2 \pi_x)$ pulse sequence has been chosen to avoid long term dephasing effects due to a miscalibration of $B_1$ amplitude and a local error in the Larmor frequency\cite{Baudin07}.

To obtain an efficient sequence, the normal $\pi/2$ rotations are replaced by magic pulses. 
We call the resulting sequence an Enhanced Magic Sandwich (EMS), since it should be more efficient in reversing the average dipolar interaction without introducing perturbations:  

\begin{equation}
[\pi_x,\pi/2_y,\pi/2_{y,0.5},-\pi/2_{y,0.5}], 
 (2 \pi_x, -2 \pi_x)_n,
 [\pi/2_y,\pi/2_{y,0.5},-\pi/2_{y,0.5}, \pi_x]
\end{equation}

\subsubsection{Larmor robust zero order M$\pi$}

M$\pi$ pulse sequences are interesting as they allow population inversion. 
The construction of such pulse is easier because less symmetries are broken by the total rotation needed which allows more flexibility in the design. 
In particular using only $\pi$ pulses of phases $\phi_1$, $\phi_2$, \textit{etc.}, the geometrical criterion for magic condition of order 0 is simply

\begin{equation}
\sum_i \tau_i e^{2 i \phi_i}=0,
\label{eq-condition_magic_pi}
\end{equation}

where $\tau_i$ is the duration of the $i$th $\pi$ pulse. 
Likewise, the criterion for Larmor shift robustness of order 0 is 

\begin{equation}
\sum_i \tau_i e^{i \phi_i}=0.
\label{eq-condition_larmor_pi}
\end{equation}

We deduce several M$\pi$ of interest for usage in experiments, for example in multiecho CP-like pulse sequences\cite{Dong07, Li08} where the dipolar interaction should not be sensitive to the pulse sequence (appart from a scaling factor). 

\begin{center}
\begin{tabular}{c|c|c|c}
Pulse sequence & Magic & $B_0$ & $B_1$\\
\hline
$\pi_x$ & $\times$ & $\times$ & $\times$\\
$2 \pi_{x},\pi_{y},\pi_{y,0.5},-\pi_{y,0.5}$ & \checkmark & \checkmark & $\times$\\
$\pi_{\pi/6},\pi_{\pi/2},\pi_{5\pi/6}$ & \checkmark  &\checkmark  &$\times$ \\
$\pi_{-\pi/6},\pi_{\pi/2},\pi_{-5\pi/6}$ & \checkmark  &$\times$  &\checkmark \\
$2\pi_{-\pi/2}, \pi_{\pi/6},\pi_{\pi/2},\pi_{5\pi/6}$ & $\times$  &\checkmark  &\checkmark \\
$2\pi_{-\pi/2}, \pi_{\pi/2}, \pi_{\pi/6},\pi_{\pi},\pi_{5\pi/6},\pi_{0}$ & \checkmark  &\checkmark  &\checkmark 
\end{tabular}
\end{center}

Magic composite $\pi$ pulses have been reported in the litterature as allowing population inversion in dipolar coupled systems\cite{Tycko84,Tycko84b}. In particular, the pulse sequence $\pi_{-\pi/6},\pi_{\pi/2},\pi_{-5\pi/6}$ have been studied by Tycko both theortically and experimentally\cite{Tycko84}. 

\section{Numerical test of the magic composite pulses}

In this section, I first explain why a numerical simulation of dipolar coupled magnetic moments is relevant to test the efficiency of magic composite pulses ususally applied in the context of quantum mechanics formalism. I will then briefly describe the simulation algorithm and present numerical evidence for the perturbation brought by the MS on the theoretical NMR evolution. Finally, I will describe the design of a numerical test for magic composite pulses and present the results on several of them. This test is a useful tool to compare the efficiency of the different magic composite pulses. 

\subsection{Equivalence between the quantum and classical description of NMR for the design of magic sequences}

The nature of the dipole-dipole interaction naturally leads to the formation of entangled states and for this reason, a classical description of NMR seems inappropriate. In liquids however, an hydrodynamic approach is suitable to describe the NMR observations because of the fast relative motion of promiscious spins and the consequent loss of quantum information\cite{Jeener02}. 
Both experimental confirmations of predictions from the classic formalism and theoretical advances now support that the classical formalism correctly describe the NMR evolution. 
In this formalism, the spin ensemble is described by the local magnetization $\vec m(\vec r,t)$ whose evolution can be described in the rotating frame by the mesoscopic Bloch-Redfield equation\cite{Jeener95} which takes into account the coupling between magnetic moments due to the distant dipolar field: 

\begin{equation}
\partial_t \vec m(\vec r,t) = \gamma \, \vec m(\vec r,t)  \times  \vec B_{d}(\vec r,t)
\label{eq-Bloch-Redfield}
\end{equation}

with

\begin{equation}
\vec B_{d} (\vec r,t) = 
\frac{\mu_0}{4 \pi} \int_{\bar V} \mathrm{d} \vec
r^{\prime}\frac 1 {|\vec r - \vec r^{\prime}|^3} (3 k_z^2 -1) \mathbf{A} \vec m (\vec r',t)
\end{equation}

where $\hat k = \frac{\vec r-\vec r^{\prime}}{|\vec r-\vec r^{\prime}|}$ and the volume of integration $\bar V$ is the total space $\mathbb{R}^3$ minus an infinitesimal spherical volume centered on the point $\vec r$. The distant dipolar field $\vec B_{d}$ is meaningful in liquids due to motional averaging of the short range dipolar interaction. This dipolar field brings a non linear and non local local term to the Bloch equation. In general, no simple solution of (\ref{eq-Bloch-Redfield}) exists for the dynamics of the magnetization. 

The Bloch-Redfield equation gives the evolution of the classical description of NMR involving coupled magnetic moments. In essence, it is truly different from the quantum description of NMR described earlier. In liquids however, this formalism describes accurately the NMR evolution. 

We are interested in using the magic composite pulses described in the previous part. Despite the different nature of the quantum (\ref{eq-Liouville}) and classical (\ref{eq-Bloch-Redfield}) NMR equations of evolution, we show in appendix \ref{appendix-Picard} that the sufficient geometrical conditions presented earlier in the quantum formalism are still valid for classical magic conditions. Furthermore, for orders 0 and 1, quantum and classical magic conditions are exactly equivalent as shown in appendix \ref{appendix-equivalence}. This equivalence between the classical and quantum magic conditions ensure that a numerical simulation of dipolar coupled magnetic moments can give relevant information about magic composite pulses used in a quantum formalism context for example for quantum information processing. Using this property we present below a numeric test of the magic composite pulses. 

\subsection{Numerical Simulation}
%copie de JLTP

To compute NMR dynamics, a general-purpose model is used based on an exact calculation of the time evolution of coupled magnetic moments on a cubic lattice. This enables to account for all the terms acting on $\vec m$ in experimental conditions in an exact manner, including diffusion, dipolar interactions, rf field, and coupling to the detection coil. However, this coarse-grained description of a continuous fluid has obvious limitations when spatial variations at the scale of the lattice constant are involved. Computations are performed in the Larmor rotating frame using a standard secular approximation. The magnetic field induced at each site by the remainder of the sample is efficiently computed by toggling between real and Fourier spaces using a half complex FFTW transform\cite{Enss99}. The time evolution is computed by integrating the Bloch-Redfield equation using a Runge-Kutta 4-6 algorithm modified to be robust against evolution discontinuities (due to rf pulses). Free evolution usually involves slow local changes and computations are quite fast on standard PC computer (Core2Duo 6700). When intense rf fields are applied, the computing load is significantly increased (e.g. by one order of magnitude for MS sequences). The computations described below have all been done on lattices of $32 \times 32 \times 32$ sites. Because of periodic boudary conditions, only $20 \times 20 \times 20$ sites were containing magnetic moments in order to account for finite size effects.

\subsection{Numerical evidence for a defect in the magic sandwich}
\label{exponential}

\begin{figure}
\begin{center}
\includegraphics[width=0.4 \textwidth]{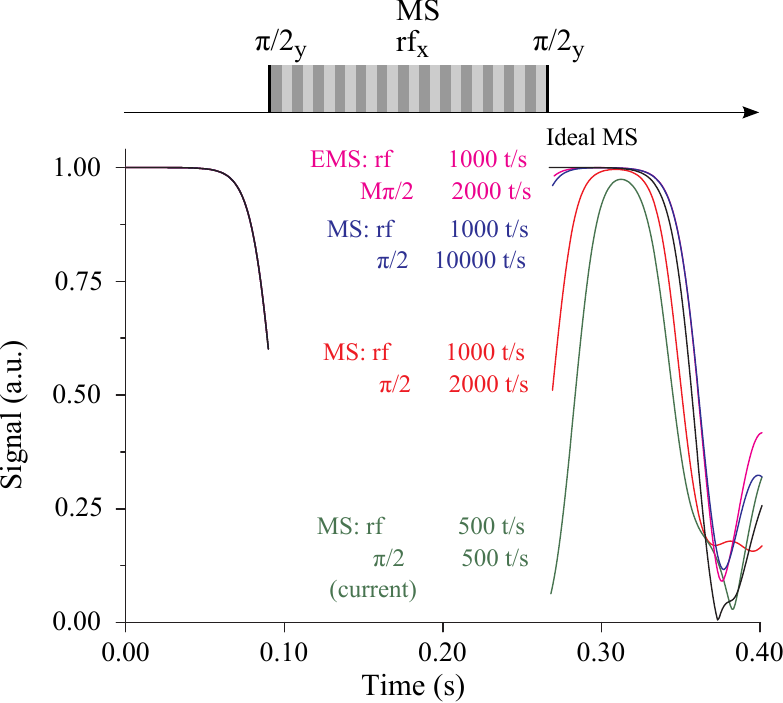}
\caption{\label{Fig6} Numerical simulations of the NMR evolution of spins submitted to MS pulse sequences. The signe is not given during the application of the MS as experimentally the rf pulse sequence saturates the detection circuit. An ideal time-reversal is given to allow a comparison of the pulse duration influence on the observed echo shape. An equivalent amplitude of 1000 turn per second during the MS filling is sufficient to reduce the perturbation brought by the filling to a negligible part of the total perturbation. The delay and amplitude of the echo is directly related to the finite duration of the initial and final $\pi/2$ rotations indicating that these are the cause for the perturbation of the evolution with regards to the ideal time-reversed evolution. 
Consequently, an EMS (see section \ref{partie-EMS}) improves the echo shape while using a lower equivalent amplitude during the initial and final $\pi/2$ rotations.}
\end{center}
\end{figure}

In the presence of long-range dipolar spin interactions, the spins in a liquid can present various collective behaviour. After a $\pi/2$ tipping pulse, the distant dipolar field (DDF) produces NMR precession instabilities which cause the signal to collapse. A model developed by Jeener and numerical simulations have shed light on magnetization dynamics\cite{Jeener99}; at large DDF, during transverse precession, unstable inhomogeneous magnetization patterns develop and grow exponentially in time at a rate $\Gamma$. These patterns arise from an initial seed of inhomogeneity following the tipping pulse. This instability produces a brutal decay of the signal. It has been shown\cite{Hayden07} that applying a magic sandwich in this liquid system allows to produce time-reversal evolution and generates a dipolar echo. However this echo of the initial signal is imperfect being both damped and delayed. Figure \ref{Fig6} represents numerical simulations of this experimental situation. Different MS have been used including an EMS using magic composite pulses for initial and final $\pi/2$ rotations. Those simulations reveal that the echo imperfection is mainly the result of the initial and final $\pi/2$ finite duration (continuous rf applied during the MS having little effect in experimental conditions). It also shows that using a magic composite pulse of order zero is equivalent in the perfect conditions of the simulation to reduce the $\pi/2$ pulse duration by one order of magnitude. 

\subsection{Design of a numerical test for magic composite pulses}

\begin{figure}
\begin{center}
\includegraphics[width=0.4 \textwidth]{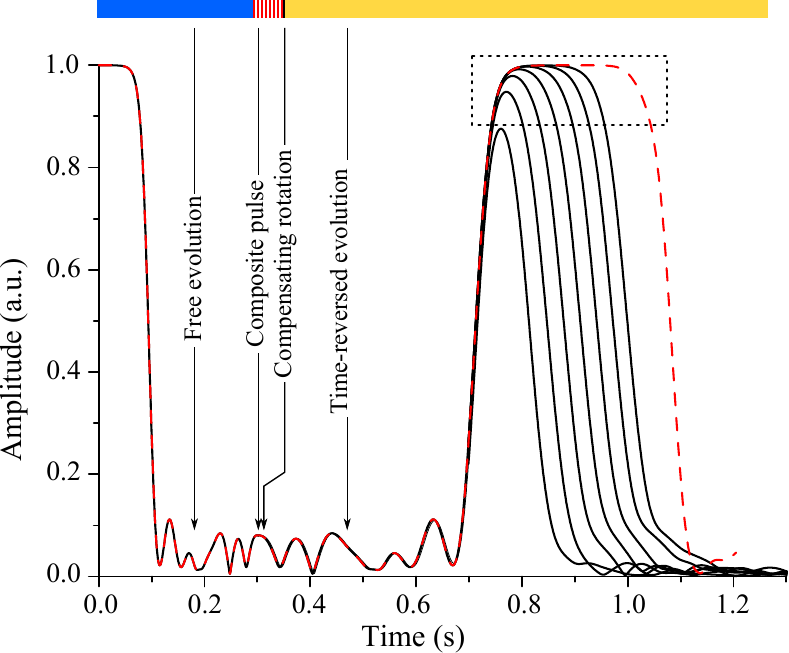}
\caption{\label{Fig1} Chronogram of the numerical test for magic composite pulse and example on a M$\\pi/2$. The different black curves are obtained for different magic composite pulse durations (durations are divided by two between two consecutive curves ranging from 0.5~s to 0.015625~ms). The deviation observed between the perfect evolution (dashed line) and the numerical tests for finite pulse sequences is due to the presence of a parasitic inhomogeneity introduced by the composite pulse. The fall time measured at a reference amplitude of 90\% is a growing affine function in the logarithm of the $B_1$ intensity. 
The dashed square indicates the valid zone where $T_{\textrm{f}}$ measurements are relevant and is defined by the conditions given in the text.}
\end{center}
\end{figure}

I consider the following numerical experiment (see chronogram in Fig. \ref{Fig1}): A free evolution followed by the pulse to be tested. Right after the pulse, an instantaneous rotation compensates its global rotation and is followed by a time-reversed evolution at half the pace of the free evolution (in order to mimic the effect of a magic sandwich). \\
The perturbation introduced by the use of the tested pulse sequence instead of a perfect instantaneous rotation is evolving in an unknown way as long as the magnetization is mainly inhomogeneous. Around the instant of the expected echo, the signal amplitude is closed to its initial value, implying that magnetization is homogeneous again\footnote{In all the presented simulations, the maps of the magnetization distribution confirm that magnetization is indeed mainly homogeneous around the instant of the echo.}. For a sufficiently small initial perturbation brought by the tested pulse, the perturbation remains small during the evolution preceeding the echo instant and, as a small perturbation, evolves according to a perturbative treatement of equation (\ref{eq-Bloch-Redfield}) in a linear manner driven by the main magnetization map\footnote{Comparisons of magnetization maps between ideal and disturbed evolutions confirm quantitatively this point.}. When those two conditions are reached (in numerical an initial free evolution time of 300 ms for 30 Hz of characteristic DDF ensures that they are), the remaining magnetization perturbation grows exponentially during the echo as it does during the intial phase of evolution. 
By measuring the collapse of the signal during the echo, we thus obtain quantitative information about the perturbation of the magnetization map brought by the tested pulse sequence.  

In figure \ref{Fig6}, the numerical is applied to the M$\pi/2$ described in \ref{ref-Mpis2} for various pulse durations. 
We note the fall time $T_{\textrm{f}}$ when the signal cross a given amplitude after the top of the echo. The relevance of this measurement is determined by several conditions: 
\begin{itemize}
\item The echo amplitude should be at least 99\% of the initial magnetization implying a mainly uniform magnetization during the echo;
\item The reference amplitude for measurement shall not be too low to ensure the perturbation brought by the pulse evolves in a linear manner. During the tests, 90\% appeared to be an acceptable value;
\item The echo does not reproduce the ideal echo shape as, in this case, the fall time is not related anymore to the perturbation introduced by the pulse. 
\end{itemize}
For this window of parameters, the fall time $T_{\textrm{f}}$ represents a global measurement of the efficiency of the pulse sequence: A benchmark of known magic pulse sequences allows to obtain numerically the order of the sequence in experimental condition. Examples given below show that this measurement of the sequence magic order is both precise and robust. This test also allows to compare the efficiency of pulse sequences having the same magic order. (This numerical test simulating the NMR dynamics during a magic sandiwch.) 

\subsubsection{Extracting informations from the measured fall time}

During the echo, magnetization beeing nearly homogeneous, the exponential development of the magnetization instability is the dominant dynamical process explaining the observed echo shape. Hence the fall time $T_{\textrm{f}}$ for a floor of 90\% is given by the following equation: 

\begin{equation}
1- \alpha \exp{\Gamma T_{\textrm{f}}}=0.9
\end{equation}

Where $\alpha$ is proportional to the size of the perturbation introduced by the tested pulse sequence. 
We thus have the relationship 

\begin{equation}
T_{\textrm{f}}= \frac{\log{0.1}-\ln{\alpha}}{\Gamma}
\end{equation}

which imply that $T_{\textrm{f}}$ is a linear function of $\log{\alpha}$. 
For a magic pulse sequence of order $n$ and duration $\tau$, the inhomogeneity size $\alpha$ is proportional to $\tau^{n+2}$, the first non canceled perturbative term in the Picard iteration of the Shroedinger equation involving $n+2$ dipolar Hamiltonians. Expliciting this dependency in the sequence order we get the relationship  
\begin{equation}
T_{\textrm{f}} \propto \ln{\alpha(\tau_0)} - (n+2) \ln{\tau/\tau_0}.
\label{eq-scale_n_Tf}
\end{equation}
By doing several numerical simulations for different pulse duration, it is possible to measure the magic order of the tested pulse. 
The duration $\tau$ is not necessarely the duration of the sequence but may be set as the duration of a typical $2\pi$ reference pulse.  
By fitting $T_{\textrm{f}}$ as an affine function of $\ln(\tau)$: 
\begin{equation}
T_{\textrm{f}} = A + B \ln(\tau),
\end{equation}
we get a measurement of the sequence's order trough coefficient $B$ and of the relative sequence efficiency trough coefficient $A$. 

\subsubsection{Completeness of the test: spin turbulence}

This numerical test using a classical formalism may seem incomplete to test the ensemble of geometrical conditions required for a magic pulse sequence of order $n$. In appendix \ref{appendix-equivalence}, we show that the magic geometrical conditions are equivalent to the magic classical conditions, the question of completeness of the test then lies in the ability of the numerical simulation to test every possibility offered by the magic geometrical conditions using a certain realization of magnetization at the instant when the pulse is applied. Although it is not possible to directly proove this point as the magnetization map is not controlled during the free evolution, it is highly unprobable that a geometrical condition would not be tested by this numerical simulation. This results from the onset of spin turbulence that magnetization undergoes before the test which corresponds to the formation of a quasi random magnetization map. Hence, this map imply many random classical conditions, and as stressed out in appendix \ref{appendix-equivalence} linear combinations of those classic conditions allow to recover the geometrical conditions. Given the huge number of random classical conditions tested, the probability of missing a magic geometrical condition is thus of measure zero. Of course symmetries of the system could default this argument, but numerical simulations are realized with an initial inhomogeneity seed chosen to break every possible symmetry of the system to avoid such problem. Consequently, the classical numerical test completely tests the magic geometrical condition and can be used safely.

\subsection{Benchmark of magic composite pulses}

The three following pulse sequences will be used as references for the magic order: 

\begin{center}
\begin{tabular}{c|c}
Pulse sequence & Magic order\\
\hline
  $\pi/2_x$ & -1 \\
  $\pi_x,\pi_y$ & 0 \\
  $\pi_x,\pi_y,\pi_{-y},\pi_{-x}$ & 1
\end{tabular}
\end{center}

\begin{figure}
\begin{center}
\includegraphics[width=0.4 \textwidth]{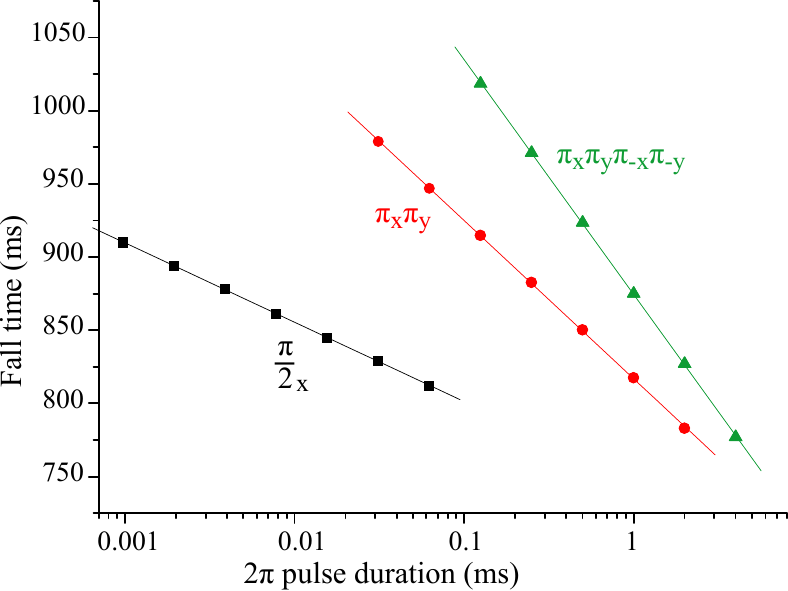}
\caption{\label{Fig2} Fall time as a function of the duration of a $2\pi$ equivalent pulse for the three reference pulse sequences $\pi/2_x$, $\pi_x,\pi_y$, and $\pi_x,\pi_y,\pi_{-x},\pi_{-y}$. The numerical test is realized for 300~ms of free evolution and a caracteristic dipolar frequency of 30~Hz. A small but relevant correction was applied prior to representation to take into account the effective time-shift on the echo instant due to the finite duration of the tested sequence. (A correction which is 3 times the duration of the tested composite pulse.)}
\end{center}
\end{figure}

The magic order $-1$ given for the pulse sequence $\pi/2_x$ exists because of the particular choice of order labelling in the average Hamiltonian theory. This order has to be interpreted as giving a perturbation scaling as $\tau^{n+2}=\tau$. 
Figure \ref{Fig2} represents the fall time $T_{\textrm{f}}$ as a function of the pulse equivalent duration. 

\begin{figure}
\begin{center}
\includegraphics[width=0.4 \textwidth]{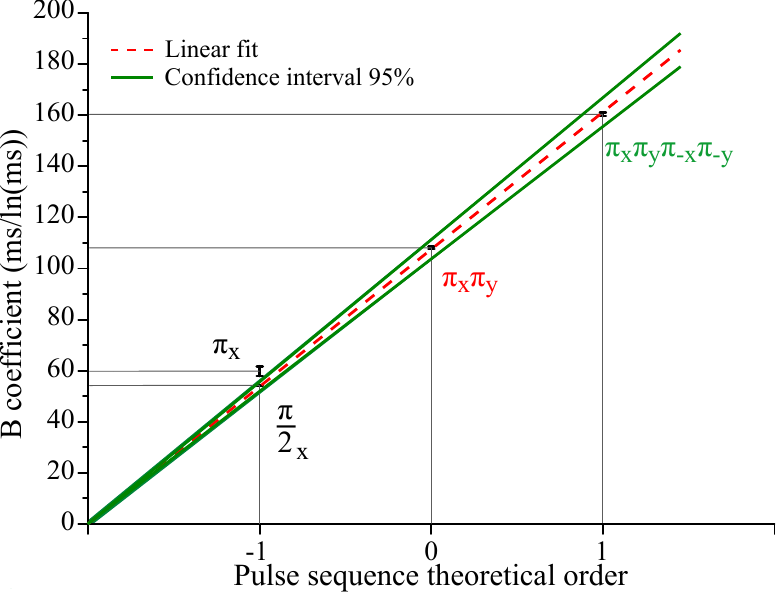}
\caption{\label{Fig3} B coefficients of the reference composite pulses measured from the data of Fig. \ref{Fig2} function of their theoretical magic order. As expected, the slopes scale linearly with $n+2$ where $n$ is the theoretical magic order of the magic composite pulse (see eq. (\ref{eq-scale_n_Tf})). A fourth pulse $\pi_x$ is also studied to infer the maximum discrepancies between the test results and the theoretical expectations. This benchmark allows to test numerically the magic order of a given composite pulse. 
}
\end{center}
\end{figure}

On Fig. \ref{Fig3},the slopes $B$ are represented as a function of the theoretical order $n$ and will serve as references to define the measured order of the magic composite pulses. 
A fourth sequence has been added on this figure: a $\pi$ rotation. This pulse presents more symmetries than the $\pi/2$ rotation an is thus of magic order 0 in duration but produces less inhomogeneity. For this reason, the measured slope of the sequence is more important than expected theoretically. This pulse serves as a guide to estimate what are the maximum possible discrepancies to be expected between theory and simulations.

Let us now turn to the following tested magic composite pulses: 

\begin{center}
\begin{tabular}{c|c|c}
Name & Pulse sequence & Order\\
\hline
$\pi/2$ & $\pi/2$ & -1\\
M$\frac{\pi}{2}$a & $\pi_x,-\pi/2_{y,1/2},\pi/2_{y,1/2},\pi/2_{y}$ & 0\\
M$\frac{\pi}{2}$b & $\pi_x,\pi/2_{y},\pi/2_{y,1/2},-\pi/2_{y,1/2}$ & 0\\
M$\frac{\pi}{2}$c & $\pi_{x,1/2},-\pi_{x,1/2},-\pi/2_{y,1/2},\pi/2_{y,1/2},\pi/2_{y}$ & 0\\
M$\frac{\pi}{2}$d & $\pi_{x,1/2},-\pi_{x,1/2},\pi/2_{y},\pi/2_{y,1/2},-\pi/2_{y,1/2}$ & 0\\
M$\frac{\pi}{2}$e & $\pi_{x,7/6},-\pi/6_y,\pi/6_y,\pi/2_y,\pi/6_y,-\pi/6_y$ & 0\\
M$\frac{\pi}{2}$ 1 &  see section \ref{Mhp1} & 1
\end{tabular}
\end{center}

\begin{figure}
\begin{center}
\includegraphics[width=0.4 \textwidth]{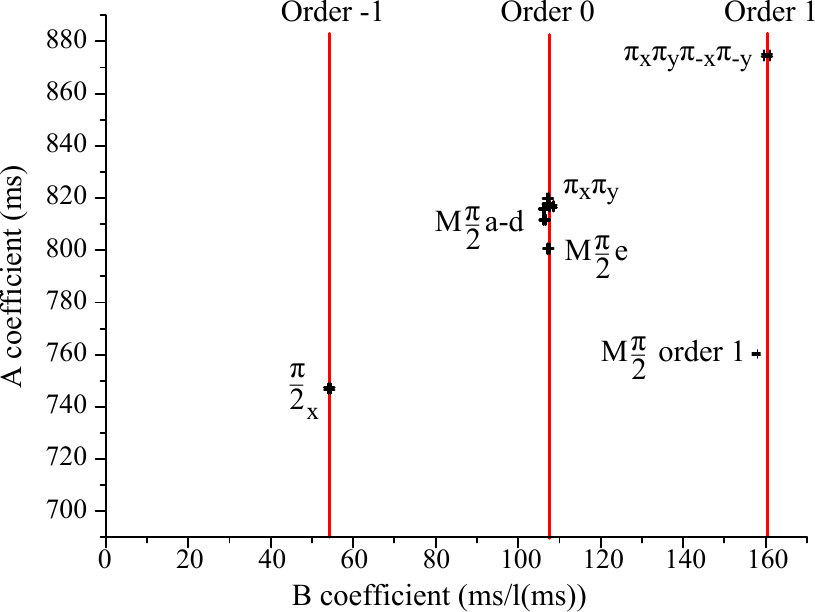}
\caption{\label{Fig4} Measurement of coefficients $A$ and $B$ of the different tested magic composite pulses. The numerical test is realized in the same conidtions as for the reference benchmark. Every simulation is realized by defining a relevant equivalent scale of duration for the $2\pi$ pulses. Vertical lines represents the expected slopes $B$ for the different theoretical orders $-1$, $0$ and $1$ extracted from Fig.~\ref{Fig3}. For a given magic order, the coefficient $A$ quantify the relative efficiency of the magic composite pulses for a given $2\pi$ reference duration. 
}
\end{center}
\end{figure}

Figure~\ref{Fig4} represents the measured parameters $A$ and $B$ for those pulse sequences. 

\subsection{Discussion of the efficiency of the different magic composite pulses}

The scale of the perturbation size brought by a magic composite pulse scales with its duration, in experimental situation, a shorter pulse will in general give a smaller perturbation. A short pulse sequence is however not always desirable for several reasons. 
For example in low temperature experiments, the amount of rf power can warm up the studied sample. Moreover, in the limit of ultra-short pulses (typically $\Omega_1/\Omega_0 \gtrsim 1/100$), rotating wave approximation is not correct anymore and corrections on the NMR dynamics have to be taken into account. Due to those corrections, the effective order of the pulse sequence may be brought back to $-1$; for example, for ultra short pulses a plain $\pi$ pulse does not produce an homogeneous rotation anymore. 

We have seen that the numerical test allows to determine reliably the pulse sequence magic order, it has been used to verify that the construction of magic composite pulses was correct. The similarity of the test with the evolution expected during a MS is of equal importance as the coefficient $A$ related to the size of the perturbation for a reference $2\pi$ pulse duration allows the comparison between different pulses in a situation close to the experimental situation: The echo observed in the numerical test corresponds to the echo that forms during the MS. 
Comparisons are relevant between pulses of the same order where the efficiency is determined by coefficent $A$ (the smaller the better), but it is also relevant when choosing between pulse sequences of different orders. Indeed, depending on the duration the pulse sequence having the highest magic order is not necesseraly the most efficient for a given equivalent $2\pi$ duration. Figure \ref{Fig1} illustrates this and show that the magic composite pulse of order 1 described in appendix \ref{appendix-Mpulse1} is not more interesting than a magic pulse sequence of order 0 in the experimental situation\cite{Hayden07} where the $2\pi$ duration is fixed between 2 and 4 ms. 
This numerical test is a method of choice to determine which, between several magic pulses candidates, is susceptible to be the most efficient magic pulse in given experimental conditions.

\begin{figure}
\begin{center}
\includegraphics[width=0.4 \textwidth]{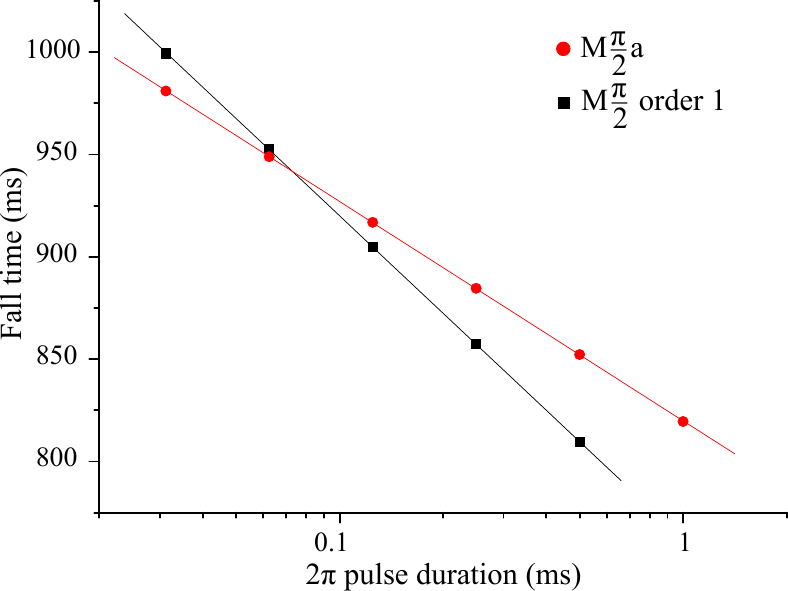}
\caption{\label{Fig5} Comparison of the fall times function of the equivalent $2\pi$ duration for the M$\pi/2$a and the M$\pi/2$ of order 1. At low rf intensity, \textit{i.e.}, long $2\pi$ equivalent duration, the zero order sequence is more efficient than the first order sequence.}
\end{center}
\end{figure}

\section{Conclusion}

In this paper, we have diagnosed that the magic sandwich is limited by the finite duration of its $\pi/2$ initial and final pulses which introduce perturbations in the time-reversal evolution. We have shown that, using the average Hamiltonian theory, it is possible to find sufficient geometrical conditions for composite $\pi/2$ which, if respected, greatly reduce the perturbation introduced when the pulse sequences are applied. We have presented a construction method for those magic composite pulses which, as opposed to the usual strategy in NMR, does not rely on the symmetries to cancel those conditions but on a direct calculation of the geometrical conditions. We then have presented a numerical test involving the numerical simulation of dipolar coupled magnetic moments on a lattice. 
After explaining how such test, though involving a classical formalism, is relevant for the study of magic composite pulses in a general context, we have explained how to obtain quantitative information about the pulse sequence efficiency. 
The results of the work presented here will be useful for solid state NMR and for the study in liquid state NMR of the complex effect of the distant dipolar field in high resolution spectrometers. Beyond NMR, those results may be of particular interest in the context of quantum information processing where dipolar couplings could be controlled using coherent control techniques similar to NMR to allow both deterministic coupling between qubits to implement quantum logic gates and decoupling of the relaxation channel between qubits and a thermal bath coupled by dipolar interactions. 

\begin{acknowledgments}
I would like to thank Pierre-Jean Nacher, Genevi\`eve tastevin, Herv\'e Desvaux and Jean Jeener for very fruitful discussions and ackowledge ANR DIPOL for funding. 
\end{acknowledgments}

\appendix 

\section{Existence of geometrical sufficient conditions for magic pulses of order $n$}
\label{appendix-geometry}
The average Hamiltonian of order $n$, $\bar H^{(n)}$, obtained by Magnus expansion\cite{Oteo00} imply a sum of $n+1$ time integrals of $n+1$ time-dependent Hamiltonians in $n$ nested commutators. Using expression (\ref{matrice_t}) for the truncated dipolar Hamiltonian, it is possible to split this expression in two parts: 
\begin{itemize}
\item A spatial part involving the coefficients $b_{ij}$ and the spin operators $\hat I_a$;
\item A geometrical part involving the matrices $\mathbf{A(t_i)}$ and the $n+1$ time integrals.
\end{itemize}
As the geometrical part depending solely on the pulse sequence and not on the sample studied, it implies that cancelling this geometrical tensor $T_{\mathbf{k_1},\mathbf{k_2},\cdots,\mathbf{k_{2n+2}}}$ ($\mathbf{k_i}=\mathbf{x},\mathbf{y} \textrm{ or } \mathbf{z}$) is a sufficient condition to ensure that $ \bar H^{(n)}=0$. 
Application to order 1 is presented in appendix \ref{appendix-Mpulse1}. 
It has to be noted that many conditions on the tensor $T$ are redundant due to symmetries of the dipolar interaction and the average Hamiltonian expression (see for example appendix \ref{appendix-equivalence}). 

\section{ A useful geometrical transformation for magic composite pulse construction}
\label{appendix-fruitful_transformation}
The computation of the quantity 

\begin{equation}
\mathbf{\bar A'} = \frac 1 {\tau} \int_{0}^{\tau} \dd t \ \mathbf{R^{-1}}(\theta(t)) \mathbf{A} \mathbf{R}(\theta(t)) 
\label{MHP_15}
\end{equation}

can be simplified by defining   

\begin{equation}
\eta (\theta) = \frac 1 {\tau} \frac 1 {\dot{\theta} (t)}
\label{MHP_16}
\end{equation}

and make a change in the integral variable: 

\begin{equation}
\mathbf{\bar A'} = \int_{\theta_m}^{\theta_M} \dd \theta \ \eta (\theta) \   \mathbf{R^{-1}}(\theta) \mathbf{A} \mathbf{R}(\theta)  
\label{MHP_17}
\end{equation}

where $\theta_M$ and $\theta_m$ are the maximum and minimum angles reached in the Bloch sphere during the time interval. 

For a usual rotation such as a $\pi_x$ for which the rotation is homogeneous and hence $\eta (\theta)=\eta_0$, this average interaction matrix is easy to compute. At the magic order 0, the weak (see appendix \ref{appendix-Mpulse1}) magic condition is fulfilled if the barycenter of the angular distribution density $\eta (\theta)$ is null. 
This representation gives a helpful picture of the sequence effect. 

\section{Construction of a magic pulse of order one}
\label{appendix-Mpulse1}

\subsection{First order magic geometrical condition}

Recalling equations (\ref{H1_limited_development}) and (\ref{eq-doubly-tilted}), the magic condition of order 1 is

\begin{equation}
\bar H_d'^{(1)} = \frac {-i}{2 \tau}\int_0^{\tau}\dd {t_1} \int_0^{t_1}  \dd {t_2}\left[ H_d'(t_1), H_d'(t_2) \right] =0.
\label{Hamiltonian_eff1_2}
\end{equation}

This condition is automatically fulfilled if the pulse sequence is time-symmetric\cite{Rhim71}, but magic composite pulses realizing a global rotation cannot possess such symmetry. Using the technique described in appendix \ref{appendix-geometry}, we derive the magic geometrical conditions of order 1: 

\begin{equation}
\int_0^{\tau}\dd {t_1} \int_0^{t_1} \dd {t_2} (\mathbf{A'}_{\mathbf{a},\mathbf{b}}(t_1)\mathbf{A'}_{\mathbf{c},\mathbf{d}}(t_2)-\mathbf{A'}_{\mathbf{a},\mathbf{b}}(t_2)\mathbf{A'}_{\mathbf{c},\mathbf{d}}(t_1))
=0
\label{Criterium_first_order}
\end{equation}

For every $(\mathbf{a},\mathbf{b},\mathbf{c},\mathbf{d})$.

\subsection{Construction of a first order magic $\pi/2$}
\label{Mhp1}

To construct the pulse sequence, we will use a recursive strategy: 
It consists in using the magic pulse of order zero and modify it until it verifies the first order too. This way, the zero order conditions are already fulfilled (which is a great simplification in the magic conditions of higher order). I make use of the symmetries in the construction process to increase the efficiency of the search by automatically fulfilling some conditions. 

As for magic composite pulses of order zero, we decompose the construction process in inhomogeneous rotations along axis $\hat x$ and $\hat y$. For an inhomogeneous rotation, the magic conditions of order zero are not well defined, but we start with a weaker condition beeing that the average matrix $\bar{\mathbf{A'}}$ is diagonal. If we call $\theta(t)$ the angle of the rotation function of time, we have the decomposition for $\mathbf{A'}(t)$: 

\begin{equation}
\mathbf{A'}(t)=
\bar{\mathbf{A'}}+\cos \theta(t) \mathbf{C} + \sin \theta(t) \mathbf{S}
\label{eq-decomposition}
\end{equation}

and  $\bar{\mathbf{A'}}$ is diagonal if and only if

\begin{equation}
\int_0^{\tau} 
\cos{\theta(t)}\dd {t}=0
\label{criterion_rotation_11}
\end{equation}

\begin{equation}
\int_0^{\tau} 
\sin{\theta(t)}\dd {t}=0.
\label{criterion_rotation_12}
\end{equation}

We will call those conditions weak magic conditions of order 0. 
Supposing those two conditions are fulfilled and introducing (\ref{eq-decomposition}) in (\ref{Criterium_first_order}), we get three geometrical conditions on the rotation to fulfill the first order magic condition for the inhomogeneous rotation: 

\begin{equation}
\int_0^{\tau} 
t \cos{\theta(t)}\dd {t}=0,
\label{criterion_rotation_7}
\end{equation}

\begin{equation}
\int_0^{\tau} 
t \sin{\theta(t)}\dd {t}=0,
\label{criterion_rotation_8}
\end{equation}

and 

\begin{equation}
\int_0^{\tau}\dd {t_1} \int_0^{t_1}  \dd {t_2} 
\sin (\theta(t_1)-\theta(t_2))=0. 
\label{criterion_rotation_4}
\end{equation}

Despite simple expressions, it is not so easy to find solutions to this set of equations. 
We define that a sequence is of the first order in a weak sense if the tensor coefficients involving at least a non diagonal term of the matrix $A'$ are all null. 
We then use the following building process: We start by building two kinds of inhomogeneous rotations that are 0-weak/1-weak: We use composite pulses realizing $\pi/2$ rotations and $2\pi$ rotations (a simple $\pi,-\pi$ pulse pair respect the given criterion for the latter case). We then bring together a  $2\pi$ inhomogeneous rotation along $\hat x$ and  a  $\pi/2$ inhomogeneous rotation along $\hat y$ having both the same total duration to obtain a magic composite pulse of order 0. As mentionned earlier, time-symmetric pulse-sequences automatically verify first order magic conditions. As we need a global rotation, we cannot directly use symmetrization here, but this symmetry argument is valid in a weak version if first order weak magic conditions are fulfilled (which is the case here). Hence we use two extra inhomogeneous rotations before the previously built pulse sequence to obtain a complete magic composite pulse of the first order. 

\begin{figure}
\begin{center}
\includegraphics[width=0.4 \textwidth]{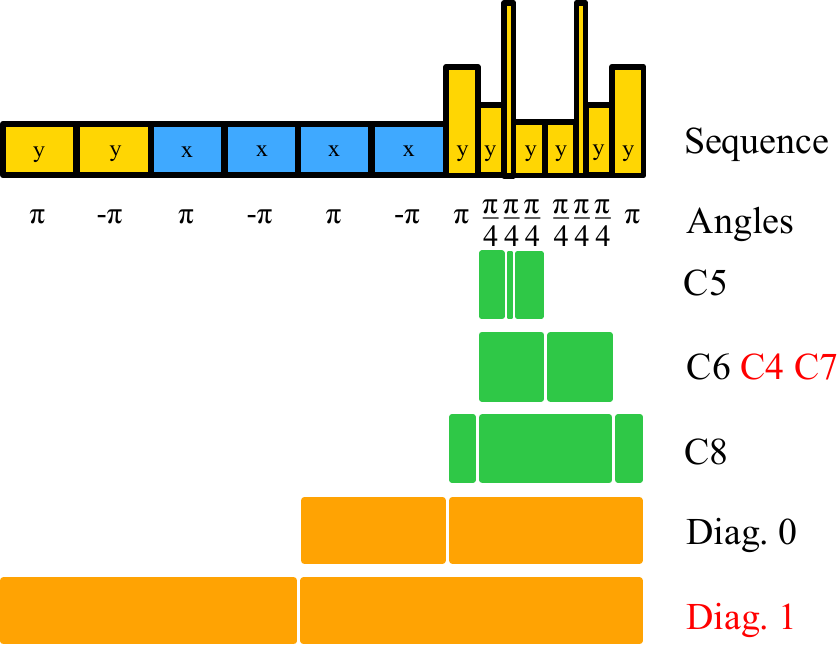}
\caption{\label{Puzzle} Schematic of the first order magic pulse construction process. Conditions in black are fulfilled by solving the corresponding equations while grey [online red] conditions are solved by using the symmetries of the pulse sequence. Diagonal conditions of order 0 and 1 are the remaining conditions implying necessarily rotations in both $\hat x$ and $\hat y$ axis allowing to turn a weak magic condition in a full magic condition. 
}
\end{center}
\end{figure}

The building process is described graphically on Fig. \ref{Puzzle}. The difficult step consists in building a 0-weak/1-weak inhomogeneous rotation. I have realized this by using $\pi/4$ pulses of different durations and using a time-symmetry to automatically fulfill two magic conditions. In practice, once the the pulse structure is determined, non linear equations on the timings are obtained from the magic conditions (\ref{criterion_rotation_11}), (\ref{criterion_rotation_12}), (\ref{criterion_rotation_7}), (\ref{criterion_rotation_8}), (\ref{criterion_rotation_4}) that take the form of zeroes of polynomials. It is clear, that without symmetry arguments, at least 5 different durations must enter in the 5 magic conditions in order to fulfill all of them. By using symmetries of the pulse sequence, I have been able to reduce this number to 3 independent parameters keeping the polynomial equations to the second degree. 

I obtain the following magic composite pulse: 
\begin{equation}
\begin{split}
[(\pi,-\pi)_{y,2.767},(\pi,-\pi)_{x,2.767},(\pi,-\pi)_{x,2.767}],[\pi_{y,1.488},
(\pi/4_{1.557},\pi/4_{1.},\pi/4_{2.557})_y,\\
(\pi/4_{2.557},\pi/4_{1.},\pi/4_{1.557})_y,\pi_{y,1.488}]
\end{split}
\end{equation}

Where the timing coefficients are zeroes of polynomials determined by conditions (\ref{criterion_rotation_11}), (\ref{criterion_rotation_12}), (\ref{criterion_rotation_7}), (\ref{criterion_rotation_8}), (\ref{criterion_rotation_4}). 

The first order magic $\pi/2$ described above is a complicated sequence. Moreover it is experimentally useless, but it allows a convincing demonstration of the geometrical construction method reported in this paper. 
This construction process also illustrates how the necessary drop of the symmetric construction strategy yields to more complex construction methods. 

\section{Picard iteration and magic conditions in the classic case}
\label{appendix-Picard}

Working with the Bloch-Redfield equation rises the problem of dealing with a non linear differential equation instead of a differential linear equation in the quantum case, hence Magnus expansion is not usable. The specificity of magic conditions \textit{i.e.} the cancelling of the average Hamiltonians for order greater than 0 allows to use the identification of the Picard iteration of the differential equation instead of Magnus expansion to obtain the magic conditions. 
Starting from a non linear differential equation of the kind: 

\begin{equation}
\partial_t Y = A(Y,t) Y,
\label{eq-Magnus_2}
\end{equation}

we can suppose that an operator $\bar A_\tau(Y)$ exists such that the solution of the non linear equation $\partial_t Y = \bar A_\tau(Y) Y$ gives a solution $Y(\tau)$ equal to the one obtained with (\ref{eq-Magnus_2}). Identification of the terms in the Picard iterations are possible if $A(Y,t)$ is a linear function of $Y$. In the case of the dipolar interaction, this requirement is verified. 
If we suppose that magic conditions up to order $n$ are fulfilled, this mean that $\bar A_\tau(Y) = \bar A_\tau^{(0)}(Y)+ O((A\tau)^{n+1})$ where $A$ quantify the size of the matrix $A(Y,t)$. Introducing this hypothesis in the Picard iteration, we get the magic condition for order 0: 

\begin{equation}
\bar A_\tau^{(0)}(Y) = k A(Y)
\label{eq-classical criterion}
\end{equation}

with
\begin{equation}
\bar A_\tau^{(0)}(Y) = \frac{1}{\tau} \int_0^\tau A(t,Y) \dd t. 
\end{equation}

And for order 1, we get 
\begin{equation}
\int_0^\tau \dd t_1 \int_0^{t_1} \dd t_2
A(t_1,A(t_2,Y)Y))-\bar A_\tau^{(0)}(\bar A_\tau^{(0)}(Y)Y) +
A(t_1,Y))A(t_2,Y)-\bar A_\tau^{(0)}(Y)\bar A_\tau^{(0)}(Y) = 
0. 
\end{equation}

The expansion is closely related to the magic conditions obtained by the Magnus expansion, but extra terms are present due to the non linearity of the equation. To stress out this similarity, we will use these non linear magic conditions on the mean-field dipolar Hamiltonian. 
It is indeed well known that the mean-field dipolar Hamiltonian

\begin{equation}
H_d = \sum_i \vec B_{di} \cdot \hat{\vec I}_i
\end{equation}

with $\vec B_{di}$ the local dipolar field which is 

\begin{equation}
\vec B_{di} = \sum_{i \neq j} b_{ij} \mathbf{A} \bra \hat{\vec I}_j  \ket
\end{equation}

used with quasi classical density matrices $\rho = \otimes_i \rho_i$ is a formalism directly equivalent to the classical formalism\cite{Jeener00}\footnote{The equivalence presented below can be found using the Bloch-redfield equation directly with the Picard iteration scheme. The mean-field formalism is presented here in order to evidence clearly the similarities between the quantum and classical conditions. A presentation which is easier as quantum and mean-field notations coincide, whereas classical notations are different. }. 
Indeed the magic condition at the order zero in this case is 

\begin{equation}
 \bar H_d^{(0)} = \frac{1}{\tau} \int_0^\tau \dd t_1 \sum_{i \neq j} b_{ij} {}^{t}\hat{\vec I}_i \mathbf{A'}(t) \bra \hat{\vec I}_j  \ket 
\end{equation}

We thus directly see the relation with the quantum condition and find that the geometrical magic condition (\ref{criterion_1}) is sufficient to ensure that the classical condition 

\begin{equation}
 \bar H_d^{(0)} = k H_d
\end{equation}

is verified. It is trivial to see that the geometrical magic condition is indeed equivalent to the classic magic condition at order 0. 

For order 1, the situation is more complexe as we have

\begin{equation}
\begin{split}
0= 
\int_0^\tau \dd t_1 \int_0^{t_1} \dd t_2
 \sum_{i \neq j} b_{ij}  \sum_{k \neq l} b_{kl} \ \ 
  {}^{t}\hat{\vec I}_i \mathbf{ A'}(t_1) \bra 
 [{}^{t}\hat{\vec I}_k \mathbf{ A'}(t_2) \bra \hat{\vec I}_l  \ket,  \hat{\vec I}_j]  \ket -
 {}^{t}\hat{\vec I}_i \mathbf{\bar A'^{(0)}} \bra 
 [{}^{t}\hat{\vec I}_k \mathbf{\bar A'^{(0)}} \bra \hat{\vec I}_l  \ket,  \hat{\vec I}_j]  \ket \\
 +
 {}^{t}\hat{\vec I}_i \mathbf{ A'}(t_1) \bra  \hat{\vec I}_j  \ket 
 {}^{t}\hat{\vec I}_k \mathbf{ A'}(t_2) \bra \hat{\vec I}_l  \ket-
 {}^{t}\hat{\vec I}_i \mathbf{\bar A'^{(0)}} \bra \hat{\vec I}_j  \ket 
 {}^{t}\hat{\vec I}_k \mathbf{\bar A'^{(0)}} \bra \hat{\vec I}_l  \ket  
 \end{split}
\label{eq-ClassicCondition_1}
\end{equation}

To obtain this expression we use the fact that $i\partial_t \bra \hat {\vec I}\ket = \bra [H_d'(t),\hat {\vec I}] \ket$. 
The expression obtained in the quantum case is the same as (\ref{eq-ClassicCondition_1}) without the brakets. We thus see that the geometrical magic criterion of order 1 is a sufficient condition to verify the magic conditions in the classical case at the order 1. In appendix \ref{appendix-equivalence}, we proove that these conditions are in fact equivalent. 
Magic conditions of higher order can be derived the same way, the only differences between quantum and classical conditions beeing those averaging operators appearing in the classic case. Those operators having no influence on the geometrical tensor, we deduce that the geometrical magic conditions are always sufficient condition both for classic and quantum magic conditions. The equivalence of classical, quantum and geometrical conditions at orders greater than one is still an open question. 

\section{Proof of the equivalence between the geometrical magic conditions and the classical magic conditions}
\label{appendix-equivalence}

To demonstrate the equivalence between the magic geometrical conditions and the classic geometrical conditions at the order one, we start from eq. (\ref{eq-ClassicCondition_1}). Let us first simplify this condition by remarking that this condition is fulfilled if it is valid for any vector on the Hilbert space. Hence we can use a braket on condition (\ref{eq-ClassicCondition_1}) to get an equivalent condition. Now let us remark the following relationship

\begin{equation}
\int_0^\tau \dd t_1 \int_0^{t_1} \dd t_2 \  \mathbf{\bar A'^{(0)}}_{\mathbf{ab}} \mathbf{\bar A'^{(0)}}_{\mathbf{cd}} = 
\int_0^\tau \dd t_1 \int_0^{t_1} \dd t_2  \ 
\{
\mathbf{A'}_{\mathbf{ab}}(t_1) \mathbf{A'}_{\mathbf{cd}}(t_2) + \mathbf{A'}_{\mathbf{ab}}(t_2) \mathbf{A'}_{\mathbf{cd}}(t_1)
\}
\end{equation}

so that we can define the tensor $T$: 

\begin{equation}
T_{\mathbf{a,b,c,d}} = 
\int_0^\tau \dd t_1 \int_0^{t_1} \dd t_2  \ 
\{
\mathbf{A'}_{\mathbf{ab}}(t_1) \mathbf{A'}_{\mathbf{cd}}(t_2) - \mathbf{A'}_{\mathbf{ab}}(t_2) \mathbf{A'}_{\mathbf{cd}}(t_1)
\}
\end{equation}

This expression allows simplifications in (\ref{eq-ClassicCondition_1})  and we get the following expression for the classical magic condition at the order one

\begin{equation}
0=\sum_{(\mathbf{a,b,c,d})\in\{x,y,z\}}
 \sum_{i \neq j} b_{ij}  \sum_{k \neq l} b_{kl} \ \ 
T_{\mathbf{a,b,c,d}} \ 
\{
\bra \hat{I}_{i,\mathbf{a}} \ket \bra [\hat{I}_{j,\mathbf{b}}, \hat{I}_{l,\mathbf{d}}]\ket \bra \hat{I}_{k,\mathbf{c}} \ket+
\bra \hat{I}_{i,\mathbf{a}} \hat{I}_{k,\mathbf{c}}\ket  \bra \hat{I}_{j,\mathbf{b}} \ket \bra \hat{I}_{l,\mathbf{d}} \ket
\} .
\end{equation}

We simplify the first part by noting that

\begin{equation}
\bra [\hat{I}_{j,\mathbf{b}}, \hat{I}_{l,\mathbf{d}}]\ket = \bra \hat{I}_{j,\mathbf{e}} \ket \varepsilon_{\mathbf{bde}} \delta_{jl}.
\end{equation}

In the contest of classical mechanics, the test vectors are limited to separable states which have the property that

\begin{equation}
\bra \hat{I}_{i,\mathbf{a}} \hat{I}_{k,\mathbf{c}}\ket = \bra \hat{I}_{i,\mathbf{a}} \ket \bra \hat{I}_{k,\mathbf{c}}\ket
\end{equation}

if $i \neq k$.
This symmetry implies that due to the structure of $T$, the term associated with this case is always null. 
If $i=k$ we have

\begin{equation}
\bra \hat{I}_{i,\mathbf{a}} \hat{I}_{k,\mathbf{c}}\ket = \bra \hat{I}_{i,\mathbf{e}} \ket \varepsilon_{\mathbf{ace}} \delta_{ik}. 
\end{equation}

The classical conditions can thus be rewritten

\begin{equation}
0=\sum_{(\mathbf{a,b,c,d})\in\{\mathbf{x},\mathbf{y},\mathbf{z}\}}
 \sum_{(i,l) \neq j} b_{ij} b_{kl} \ \ 
T_{\mathbf{a,b,c,d}} \ \ 
\{
\varepsilon_{\mathbf{bde}} \delta_{jl} \bra \hat{I}_{i,\mathbf{a}} \ket \bra \hat{I}_{j,\mathbf{e}} \ket  \bra \hat{I}_{k,\mathbf{c}} \ket+
\varepsilon_{\mathbf{ace}} \delta_{ik} \bra \hat{I}_{i,\mathbf{e}} \ket \bra \hat{I}_{j,\mathbf{b}} \ket \bra \hat{I}_{l,\mathbf{d}} \ket
\}
\end{equation}

Simple permutation of indexes using the symmetry $b_{ij}=b_{ji}$ allows to write the condition in the form

\begin{equation}
0=
\sum_{(i,l) \neq j} b_{ij}  b_{jl} 
\sum_{(\mathbf{a,e,d})\in\{x,y,z\}}
Q_{\mathbf{a,e,d}}
\bra \hat{I}_{i,\mathbf{a}} \ket \bra \hat{I}_{j,\mathbf{e}} \ket  \bra \hat{I}_{l,\mathbf{d}} \ket
\label{eq-Simplifie_magic_1}
\end{equation}

where $Q_{\mathbf{a,e,d}} = T_{\mathbf{a,b,c,d}} \varepsilon_{\mathbf{bcd}}$ which has the property 
\begin{equation}
Q_{\mathbf{a,e,d}}=Q_{\mathbf{d,e,a}}.
\label{eq-tensor-symmetry}
\end{equation} 
The equation (\ref{eq-Simplifie_magic_1}) gives the classical magic conditions of order one while the equation $Q_{\mathbf{a,e,d}}=0$ gives the geometrical magic conditions. 
It is trivial to see that the geometrical conditions imply that the classic conditions is verified. The opposite is more complex but feasable by supposing a sample of 3 spins with the different separable states. 

Let's start with a sample prepared in the state $|z_+,z_+,z_+\ket$ (the order $n$ in braket notations corresponds to the spin number $n$). 
The classic condition in this particular case is 

\begin{equation}
0=
\sum_{(i,l) \neq j} b_{ij}  b_{jl} 
Q_{\mathbf{z,z,z}}.
\end{equation}

If the positions of the spins are correctly chosen the spatial part is non zero which imply that $Q_{\mathbf{z,z,z}}=0$. Consequently by changing the test vector we also get two other geometrical conditions: $Q_{\mathbf{x,x,x}}=Q_{\mathbf{y,y,y}}=0$. 

We now use a sample prepared in the state $|x_+,x_+,y_+\ket$ and compute the classic condition. After some algebra we get

\begin{equation}
0=\alpha_1 Q_{\mathbf{x,y,x}} + \alpha_2 (Q_{\mathbf{x,x,y}}+Q_{\mathbf{y,x,x}})+ \alpha_3 Q_{\mathbf{y,x,y}}
+\textrm{conditions already verified}
\end{equation}

with 

\begin{equation}
\left \{
\begin{array}{c @{=} c}
    \alpha_1 & 2 b_{13} b_ {23} + b_{13}^2 + b_{23}^2\\
    \alpha_2 & b_{12} b_ {23} + b_{12} b_ {13}\\
    \alpha_3 & b_{13}^2 + b_{23}^2.
\end{array}
\right.
\end{equation}

This condition must be met for any choice of position. The difference in the change of $\alpha_1$ and $\alpha_2$ with the choice of the spin positions imply that the previous condition is valid for any choice of $\alpha_1$ and $\alpha_2$. We deduce the geometrical conditions $Q_{\mathbf{x,y,x}}=0$ and $Q_{\mathbf{x,x,y}}=0$ (remembering that the sufficient geometrical conditions has the symmetry (\ref{eq-tensor-symmetry})). By choosing other directions in the test vector we get the geometrical conditions for every choice in the tensor $Q$ with 2 directions. 

Finally, we choose the test vector $|x_+,y_+,z_+\ket$ and get the condition: 

\begin{equation}
0=\beta_1 (Q_{\mathbf{x,y,z}}+Q_{\mathbf{z,y,x}}) 
+\beta_2 (Q_{\mathbf{y,z,x}}+Q_{\mathbf{x,z,y}}) 
+\beta_3 (Q_{\mathbf{z,x,y}}+Q_{\mathbf{y,x,z}}) 
+\textrm{conditions already verified}
\end{equation}

with 

\begin{equation}
\left \{
\begin{array}{c @{=} c}
    \beta_1 & b_{12} b_ {23} \\
    \beta_2 & b_{13} b_ {23} \\
    \beta_3 & b_{12} b_ {13}.
\end{array}
\right.
\end{equation} 

The same reasoning applies here: The position of the 3 spins can be chosen arbitrarily so that the coefficients $\beta$ are independent. We know already that the conditions with less than 3 coordinates must be canceled in order to have a magic sequence so that we get finally all geometrical conditions (using the symmetry of equation (\ref{eq-tensor-symmetry}): $Q_{\mathbf{a,e,c}}=0$ for any choice of $\mathbf{a}$, $\mathbf{e}$ and $\mathbf{c}$. 
We thus proved the equivalence between the geometrical and classical conditions. 
The equivalence with the quantum conditions  is likewise proovable. 
Extensive analysis using Mathematica show that they are only 6 independent first order geometrical magic conditions. 
By using eq. (\ref{matrice_A_prime2}), it is easy to show that quantum and classical conditions can be written as linear set of equations involving the geometrical conditions. LU decomposition allows to show that the number of independent geometrical conditions is 6 both in quantum and classical case. 
In the general case of order $n \geq 2$ (a case not considered in practice in this paper), the equivalence between classical, quantum and geometrical magic conditions still remain to be proven. 

%Pour la bibliographie sur ArXiv


\begin{thebibliography}{10}

\bibitem{Abragam61}
A.~Abragam.
\newblock {\em The Principles of Nuclear Magnetism}.
\newblock Oxford at the Clarendon Press, 1961.

\bibitem{Barenco95}
Adriano Barenco, David Deutsch, Artur Ekert, and Richard Jozsa.
\newblock Conditional quantum dynamics and logic gates.
\newblock {\em Phys. Rev. Lett.}, 74:4083--4086, May 1995.

\bibitem{Krojanski06}
Hans~Georg Krojanski and Dieter Suter.
\newblock Reduced decoherence in large quantum registers.
\newblock {\em Phys. Rev. Lett.}, 97:150503, Oct 2006.

\bibitem{Rhim71}
W-K. Rhim, A.~Pines, and J.S. Waugh.
\newblock Time-reversal experiments in dipolar-coupled spin systems.
\newblock {\em Phys. Rev. B}, 3(3):684--696, 1971.

\bibitem{Hayden07}
M.E. Hayden, E.~Baudin, G.~Tastevin, and P.J. Nacher.
\newblock Nmr time reversal as a probe of incipient turbulent spin dynamics.
\newblock {\em Phys. Rev. Lett.}, 99(13):137602, Sep 2007.

\bibitem{Jeener02}
J.~Jeener.
\newblock {\em Encyclopedia of Nuclear Magnetic Resonance, Volume 9: Advances
  in NMR}.
\newblock John Wiley \& Sons, 2002.

\bibitem{Nacher00}
P.-J. Nacher, G.~Tastevin, B.~Villard, N.~Piegay, F.~Marion, and K.~Sauer.
\newblock Nmr instabilities in spin-polarized liquids : He3, he3-he4 mixtures
  and xe129.
\newblock {\em J. Low Temp. Phys.}, 121(5-6):743--748, Dec 2000.

\bibitem{Sauer01}
K.~L. Sauer, F.~Marion, P.~J. Nacher, and G.~Tastevin.
\newblock Nmr instabilities and spectral clustering in laser-polarized liquid
  xenon.
\newblock {\em Phys. Rev. B}, 63:184427, Apr 2001.

\bibitem{Lin00}
Y.-Y. Lin, N.~Lisitza, S.~Ahn, and W.S. Warren.
\newblock Resurrection of crushed magnetization and chaotic dynamics in
  solution nmr spectroscopy.
\newblock {\em Science}, 290(5489):118--121, 2000.

\bibitem{Levitt86}
M.H. Levitt.
\newblock Composite pulses.
\newblock {\em Prog. Nucl. Mag. Res. Sp.}, 18(2):61 -- 122, 1986.

\bibitem{Waugh07}
John~S. Waugh.
\newblock {\em Average Hamiltonian Theory}.
\newblock John Wiley \& Sons, Ltd, 2007.

\bibitem{Maricq82}
M.~Matti Maricq.
\newblock Application of average hamiltonian theory to the nmr of solids.
\newblock {\em Phys. Rev. B}, 25(11):6622--6632, Jun 1982.

\bibitem{Magnus54}
W.~Magnus.
\newblock On the exponential solution of differential equations for a linear
  operator.
\newblock {\em Commun. Pure Appl. Math.}, 7:649, 1954.

\bibitem{Wilcox67}
R.M. Wilcox.
\newblock Exponential operators and parameter differentiation in quantum
  physics.
\newblock {\em J. Math. Phys.}, 8(4):962--983, 1967.

\bibitem{Haeberlen68}
U.~Haeberlen and J.~S. Waugh.
\newblock Coherent averaging effects in magnetic resonance.
\newblock {\em Phys. Rev.}, 175(2):453--467, Nov 1968.

\bibitem{Oteo00}
J.~A. Oteo and J.~Ros.
\newblock From time-ordered products to magnus expansion.
\newblock {\em J. Math. Phys.}, 41(5):3268--3277, 2000.

\bibitem{Casas07}
Fernando Casas.
\newblock Sufficient conditions for the convergence of the magnus expansion.
\newblock {\em Journal of Physics A: Mathematical and Theoretical},
  40(50):15001, 2007.

\bibitem{Blanes09}
S.~Blanes, F.~Casas, J.A. Oteo, and J.~Ros.
\newblock The magnus expansion and some of its applications.
\newblock {\em Physics Reports}, 470(5–6):151 -- 238, 2009.

\bibitem{Rhim73}
W.-K. Rhim, D.D. Elleman, and R.W. Vaughan.
\newblock Analysis of multiple pulse nmr in solids.
\newblock {\em J. Chem. Phys.}, 59(7):3740--3750, 1973.

\bibitem{Rhim74}
W.-K. Rhim, D.D. Elleman, L.B. Schreiber, and R.W. Vaughan.
\newblock Analysis of multiple pulse nmr in solids. ii.
\newblock {\em J. Chem. Phys.}, 60(11):4595--4605, 1974.

\bibitem{Redfield55}
A.G. Redfield.
\newblock Nuclear magnetic resonance saturation and rotary saturation in
  solids.
\newblock {\em Phys. Rev.}, 98(6):1787--1809, 1955.

\bibitem{Baudin07}
E.~Baudin, M.E. Hayden, G.~Tastevin, and P.-J. Nacher.
\newblock Nonlinear nmr dynamics in hyperpolarized liquid 3he.
\newblock {\em C. R. Chim.}, 11(4-5):560--567, Apr 2007.

\bibitem{Dong07}
Dale Li, A.~E. Dementyev, Yanqun Dong, R.~G. Ramos, and S.~E. Barrett.
\newblock Generating unexpected spin echoes in dipolar solids with pi pulses.
\newblock {\em Phys. Rev. Lett.}, 98(19):190401, May 2007.

\bibitem{Li08}
Dale Li, Yanqun Dong, R.~G. Ramos, J.~D. Murray, K.~MacLean, A.~E. Dementyev,
  and S.~E. Barrett.
\newblock Intrinsic origin of spin echoes in dipolar solids generated by strong
  $\pi$ pulses.
\newblock {\em Phys. Rev. B}, 77:214306, Jun 2008.

\bibitem{Tycko84}
R.~Tycko, E.~Schneider, and A.~Pines.
\newblock Broadband population inversion in solid state nmr.
\newblock {\em J. Chem. Phys.}, 81(2):680--688, 1984.

\bibitem{Tycko84b}
R.~Tycko and A.~Pines.
\newblock Iterative schemes for broad-band and narrow-band population inversion
  in nmr.
\newblock {\em Chem. Phys. Lett.}, 111(4-5):462 -- 467, 1984.

\bibitem{Jeener95}
J.~Jeener, A.~Vlassenbroeck, and P.~Broekaert.
\newblock Unified derivation of the dipolar field and relaxation terms in the
  bloch-redfield equations of liquid nmr.
\newblock {\em J. Chem. Phys.}, 103(4):1309--1333, 1995.

\bibitem{Enss99}
T.~Enss, S.~Ahn, and S.~Warren Warren.
\newblock Visualizing the dipolar field in solution nmr and mr imaging:
  three-dimensional structure simulations.
\newblock {\em Chem. Phys. Lett.}, 305(1-2):101--108, May 1999.

\bibitem{Jeener99}
J.~Jeener.
\newblock Dynamical effects of the dipolar field inhomogeneities in
  high-resolution nmr: Spectral clustering and instabilities.
\newblock {\em Phys. Rev. Lett.}, 82(8):1772--1775, 1999.

\bibitem{Jeener00}
J.~Jeener.
\newblock Equivalence between the ``classical'' and the ``warren'' approaches
  for the effects of long range dipolar couplings in liquid nuclear magnetic
  resonance.
\newblock {\em The Journal of Chemical Physics}, 112(11):5091--5094, 2000.

\end{thebibliography}
\end{document}